# LLMRS: Unlocking Potentials of LLM-Based Recommender Systems for Software Purchase

Angela John*
Saarland University
ajohn@cs.uni-saarland.de

Theophilus Aidoo*
Saarland University
taidoo@cs.uni-saarland.de

Hamayoon Behmanush*
Saarland University
Habe00007@uni-saarland.de

Irem B. Gunduz**
Saarland University
Irgu00001@uni-saarland.de

Hewan Shrestha**
Saarland University
Hesh00001@uni-saarland.de

Maxx Richard Rahman
Saarland University
m.rahman@iss.uni-saarland.de

Wolfgang Maaß[†]
Saarland University
wolfgang.maass@iss.uni-saarland.de

*Equal Contribution (~first)
**Equal Contribution (~second)
[†]Corresponding Author

*Submission Type: Completed Full Research Paper*

## Abstract

Recommendation systems are ubiquitous, from Spotify playlist suggestions to Amazon product suggestions. Nevertheless, depending on the methodology or the dataset, these systems typically fail to capture user preferences and generate general recommendations. Recent advancements in Large Language Models (LLM) offer promising results for analyzing user queries. However, employing these models to capture user preferences and efficiency remains an open question. In this paper, we propose LLMRS, an LLM-based zero-shot recommender system where we employ pre-trained LLM to encode user reviews into a review score and generate user-tailored recommendations. We experimented with LLMRS on a real-world dataset, the Amazon product reviews, for software purchase use cases. The results show that LLMRS outperforms the ranking-based baseline model while successfully capturing meaningful information from product reviews, thereby providing more reliable recommendations.

## Introduction

Recommender systems are used as a bridge between the customer and product catalogs. These systems employ various filtering methodologies to select the top product from the product pool based on a customer's needs and preferences. These methodologies can be summarized into three categories: collaborative filtering, content-based filtering, and hybrid filtering. One widely used filtering approach is filtering the products that new users may like based on the similarity of previous users' profiles and reviews of products, which is called collaborative filtering [1]. This approach requires an extensive dataset rich with different user profiles, preferences, and reviews. Another approach is content-based filtering, which makes recommendations based on product-to-product similarity. This approach, on the other hand, depends on the user's explicit feedback or reactions. There are also hybrid approaches, which employ both content-based and collaborative filtering [1].

Recently, recommender systems started to build upon complex algorithms to capture users' reviews better than simple filtering-based approaches. However, these approaches depend on big datasets and can be computationally expensive. Pre-trained Large Language Models (LLMs), on the other hand, show promise for analyzing user queries and do not need a lot of computing power. However, these models have limited capacity to capture the meaning of sentences, even hallucinate, and require precise prompting. Therefore, how LLM can be used to incorporate user reviews into the recommendation system and how to choose the most suitable LLM remain an open challenge.

**Related Works**

Recommender systems have dramatically improved recently. To enhance the precision and effectiveness of suggestions, researchers have examined a variety of techniques. These innovations have progressed the sector and made it possible for people to receive more pertinent recommendations that meet their needs and interests. Future research in this field looks to bring up some intriguing new advances.

Widely used approaches for providing individualized suggestions to people include content-based, explainable, and collaborative recommendation systems. Hariadi and Nurjanah [2] developed a hybrid attribute- and personality-based book recommendation system to address over-specialized suggestions and cold-start circumstances. To make individualized recommendations, they made use of user profiles, similarity measures, and book rating projections. Wang et al. [3] proposed a content-based recommender system for computer science papers that focused on conferences and academic publications. To deal with the cold start problem and the need for human changes when training datasets changed, they used content-based filtering with chi-squared and softmax regression.

Kido et al. [4] focused on user-written assessments and provided clear recommendations using the EFM and MTER approaches, but despite using reviews, this study had a shortcoming in effectively filtering reviews for performance. We address this by using LLM models to generate positivity and negativity scores for each review. To solve concerns with hallucinatory outcomes and information gaps, Yang et al. [5] introduced the PALR framework, including historical interactions and language model-based algorithms for tailored suggestions. Gao et al. [6]

introduced the T-REC framework and also demonstrated how well it performed in comparison to other systems. It is a language model-based interactive and explicable recommender system that makes use of user-written reviews and interactions. Markov Decision Processes (MDP) are used as a framework to make recommendations to maximize customer happiness [7]. Shoja and Tabrizi [8] proposed a method that uses Latent Dirichlet Allocation (LDA) for customer review analysis and matrix factorization for rating-based suggestions. Within recommender systems, the use of ranking processes is of utmost importance.

By incorporating rankings, recommender systems are given the ability to respond to a wide range of user needs, improving the overall user experience and encouraging user involvement. Alhijawi et al. [12] proposed item popularity and predicted ranking as a Multi-Factor Ranking (MF-R) for collaborative-based recommender systems. The MF-R improves the quality and thoroughness of its recommendations by conducting a systematic search for proposals that meet the criteria of correctness, originality, diversity, and coverage. By employing this technique, it is ensured that the recommendations are not only precise but also varied and inclusive, taking a wider range of user preferences into account. LLMs have great potential for enhancing recommender systems. According to Hou et al. [13], a method that frames LLMs as conditional ranking tasks enables their use for recommendation procedures. They use LLMs to create personalized rankings based on prior encounters, which boosts the effectiveness of recommender systems, although this model suffers from popularity bias which we address in our proposed model by scaling positivity score to match the number of reviews and generating ratings based on it.

Despite tremendous advancements, there are still problems to be solved in the industry. Ramni et al. [9] discussed computational difficulty and lack of specificity regarding dataset choice in earlier methods. Another study raised concerns about violating data privacy rules because it lacked an established dataset [3]. Gao et al.'s method for explainable recommendations [6] also has a long calculation time drawback, this paper uses LLM but it ignores reviews which is believed to carry much information and instead uses ratings and in our proposed method, we rely on reviews for recommendation. We find the work of Shivangi et el [17], to be very similar to ours in that it uses an LLM approach to rank recommendations to users. However, it considers the previous interaction of a customer to produce personalized recommendations. In our approach, however, we rely on review text to make recommendations to users.

In this paper, we propose LLMRS, an LLM-based zero-shot recommender system, to overcome these challenges and unlock the opportunities of LLMs to capture user preferences and generate user-tailored recommendations.

**Proposed Pipeline: LLMRS**

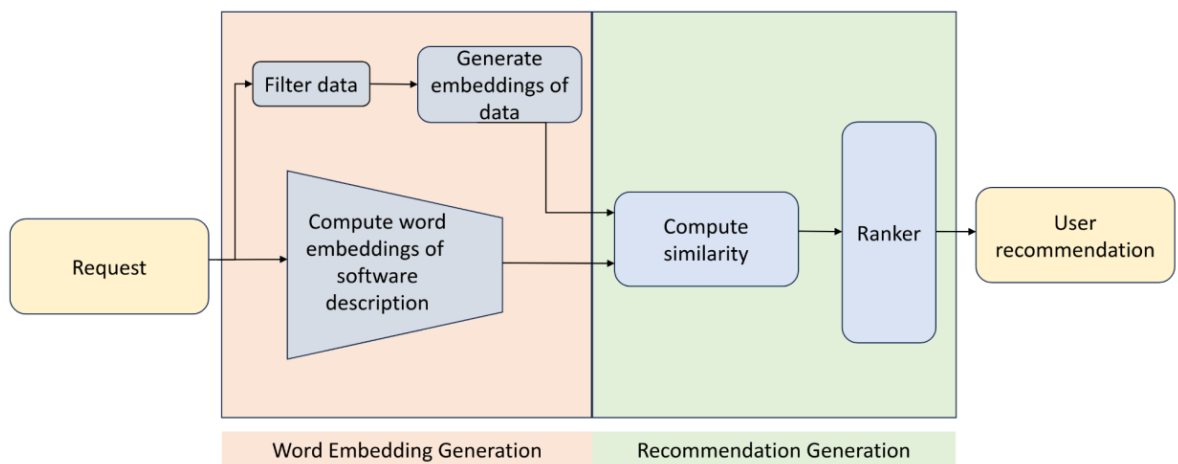

Figure 1: LLMRS workflow

The proposed method follows the approach illustrated in Figure 1. It consists of two main components, namely, I) Word Embedding Generation, and II) Recommendation Generation.

**Word Embedding Generation**

The input to our model is a request query text from a user that is formatted as a string. The requested query also includes price, maintenance cost, license fees, and implementation cost. The total data is filtered with the monetary values from the request before the computation of the similarity score, as a measure to reduce the number of computations.

We compute a numerical embedding vector for both to compare it to previous knowledge from the database. We used the Masked and Permuted Pre-Training for Language Understanding (MPNet), a transformer-based model, to compute our embedding vectors. MPNet is a pre-trained LLM model formed from Masked Language Modeling (MLM) in BERT and permutation language from XLNet, combining the strengths of the two transformers [10]. It reduces the disparity between pre-training and the anticipated tokens by allowing the model to observe auxiliary position information. To speed up model performance at inference, we pre-computed the embeddings of our dataset and saved them as a file for which the query embedding can be compared. Given that t is the time and $x$ is the sentence, $c$ denotes the number of non-predicted tokens, $z$ is the permutation, $x_{\{z<t\}}$ tokens predict the currently predicted tokens, $x_{\{z_t\}}$ and $M$ is the mask of tokens. The MPNet training objective is given by:

$$\mathbb{E}_{z \in Z_n} \sum_{t=c+1}^{n} \log P(x_{z_t} | x_{z<t}, M_{z>c}; \theta)$$

**Recommendation Generation**

The generated embeddings from the previous stage are passed to a similarity function to compute the degree of similarity between each pair of embedding vectors. We used cosine as the similarity function, where a higher value means the vectors are very similar and a low value means the vectors are dissimilar. This is given by:

$$cos\,(\theta) = \frac{A \cdot B_i}{\|A\|\|B_i\|}$$

where $A$ is the embedding vector of the request text, $B_i$ is the embedding vector of the $i^{th}$ index of our dataset, and this was done with overall samples of software products. The computed similarities are used to compare the requested text (A) and each software description ($B_i$)in our dataset this pre-select the software products that are similar to the requested text. This is done to ensure that at least the textual description of the requested text is related to the description to the preselected software.

The computed similarities are passed to the Ranker. We used K-means to cluster ratings generated through the Term Frequency-Inverse Document Frequency (TF-IDF). The TF-IDF matrix is a measure of word importance in the review texts; frequent words are given higher weights than less frequently occurring words. Using the K-means clustering algorithm, we obtained five clusters from the reviews' TF-IDF matrix using the K-means clustering algorithm. We used a zero-shot classification LLM to generate positive and negative scores for each review text. Then the count of each label in all clusters was computed and used for the cluster rating. The cluster rate, denoted as $Y$, with positive count $x_p$ and negative count $x_n$, can be expressed as:

$$Y = x_p / (x_p + x_n)$$

We assigned ratings based on the cluster with the highest positive sentiments relative to others. That is, the cluster with the highest value is rated as 5, and the cluster with the lowest positive score is rated as 1. The resulting ratings are consistent, as each review is attached to a review with an appropriate weight of positivity. In our study, we utilized a zero-shot approach using a pre-trained model, Bart [16], an aspect-level sentiment analysis model. This model is used to generate the negative and positive sentiment scores per review. These scores were then used to generate rank scores for recommending products to users in the recommendation system. Assuming the rank score is denoted as $R$, the negative score as $N$, the positive score as $P$, and the total number of reviews for the software as $S$, our rank is computed using the following formula:

$$R = (P - N) * S$$

$S$ is a weighting parameter for the difference between the positive and negative sentiment measures.

## Experiments

### Datasets

We used the software category of the Amazon Reviews Dataset, which has product reviews and metadata with 17,424 entries. The dataset contains the following features as columns: category, description, brand, feature information, sales rank, number of views, products also viewed with corresponding software, price, and date information per software product. We also simulated maintenance costs, licensing fees, and implementation costs from the price feature and added them as additional features. The dataset we used for this work did not contain information on licensing fees, implementation costs, and maintenance costs which are typical information provided for every software on sale. We simulated licensing fees as 80% of the minimum price per software category, implementation costs as 50% of the price per software, and maintenance costs as 1% of the price. We assume licensing and implementation fees are paid once, and maintenance fees are paid as monthly fees; also, all monetary values are in 100s of US Dollars($). The user reviews contain 400,723 reviews, including verified labels, review IDs, reviewer names, review texts, and summaries. After filtering out unique software products, we ended up with 3605 unique software products with a price range of between 0 and 3175. The most reviewed software had 8990 reviews, with the least reviewed having 1 review. On average, each software had 39 reviews, as can be seen in Table 1.

|  | Price | License fee | Implementation cost | Maintenance cost | Positive score | Negative score | Number of reviews |
|---|---|---|---|---|---|---|---|
| **Count** | 3605 | 3605 | 3605 | 3605 | 3605 | 3605 | 3605 |
| **mean** | 48.73 | 0.99 | 24.37 | 4.87 | 24.40 | 13.62 | 39.33 |
| **std** | 132.05 | 1.52 | 66.02 | 13.20 | 156.23 | 58.76 | 222.94 |
| **min** | 0.00 | 0.00 | 0.00 | 0.00 | 0.00 | 0.00 | 1.00 |
| **25%** | 10.98 | 0.00 | 5.49 | 1.10 | 0.75 | 0.87 | 1.00 |
| **50%** | 19.99 | 0.01 | 9.99 | 1.99 | 1.87 | 1.74 | 3.00 |
| **75%** | 43.90 | 1.07 | 21.95 | 4.39 | 7.18 | 5.79 | 13.00 |
| **max** | 3175.00 | 11.96 | 1587.50 | 317.50 | 6559.04 | 1778.95 | 8990.00 |

Table 1: Data descriptive statistics

We did not use features of the dataset that relate to personal information, such as the reviewer's name and location, as it was not our intention to use such information in this project. However, in the future, we could consider location data for a geographical customization of recommendations.

**Baseline**

Recommender systems can utilize historical information such as ratings and reviews to make recommendations. However, due to the challenges of reconciling reviews with corresponding ratings, most of the traditional systems ignore this information. Thus, suffer from the cold start problem when there is insufficient correct review-to-rating information. Generally, it is expected that positive reviews are associated with high ratings and negative reviews are associated with low ratings. This, however, is not how the Amazon dataset is actually represented; Figure 2 illustrates this. At the lower end, 15493 entries are rated as 1 when the reviews are positive, while 23110 negative reviews are rated as 5 stars, but the review scores have a higher negative score. Therefore, in our LLMRS model, we regarded reviews as a source of information for generating consistent ratings. With our baseline, we use the average of the provided ratings to recommend the software products.

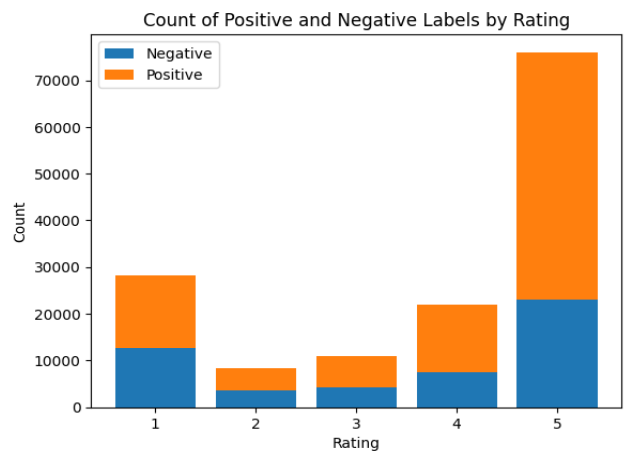

Figure 2: Overview of positive and negative labeled products

To develop our LLMRS model, we experimented with four models: Term Frequency-Inverse Document Frequency (TF-IDF), which is the most commonly used algorithm to transform text into meaningful numbers used in machine learning tasks [11], Masked and Permuted Pre-Training for Language Understanding (MPNet) [10], A Lite BERT for Self-Supervised Learning of Language Representations (Albert) [14], and MiniLM [15], to generate embeddings based on cosine similarity, which measures how similar or the same two non-zero vectors defined in an inner product space are, and ranking score recommendation software products. From the experiment, MPNet performed  better than all others, so we implemented it with our LLMRS model.

## Results

We benchmarked LLMRS against a baseline model, which utilizes the cosine similarity of the query text and item description and ranks using ratings.

**Query Text: "HR program for managing employee records on Windows platform"**

| Description | Price | Licenc. Fee | Implem Fee | Main. Fee | Rank Score |
|---|---|---|---|---|---|
| anytime organizer every tool need or… | $3,999.00 | $0.80 | $1,999.50 | $399.90 | 591.79 |
| the right tool job . write format audit resu… | $3,995.00 | $0.80 | $1,997.50 | $399.50 | 153.97 |
| winway resume deluxe includes new 8220 save… | $1,999.00 | $0.80 | $999.50 | $199.90 | 99.81 |
| microsoft office small business management edi… | $12,990.00 | $0.80 | $6,495.00 | $1,299.00 | 11.82 |
| anytime organizer every tool need organize p… | $1.00 | $0.00 | $0.50 | $0.10 | 2.43 |

Figure 3: Top five recommendations match using LLMRS using the query.

| Description | Price | Licenc. Fee | Implem Fee | Main. Fee | Avg Rating |
|---|---|---|---|---|---|
| staff file perfect solution today manager nee… | $21,780.00 | $0.80 | $10,890.00 | $2,178.00 | 5.00 |
| orgchart professional delivers world-class org… | $9,999.00 | $0.80 | $4,999.50 | $999.90 | 5.00 |
| microsoft office small business management edi… | $12,990.00 | $0.80 | $6,495.00 | $1,299.00 | 5.00 |
| the right tool job . write format audit resu… | $3,995.00 | $0.80 | $1,997.50 | $399.50 | 4.07 |
| anytime organizer every tool need or… | $3,999.00 | $0.80 | $1,999.50 | $399.90 | 3.56 |

Figure 4: Top five recommendations from baseline model using the query.

The top five recommendations from our model (Figure 3) and baseline model (Figure 4) return recommendations related to the query text. The baseline model was ranked based on the average ratings of each software product. The average rating is the mean of all ratings that previous users have given to the software. Our model, on the other hand, used the proposed ranker algorithm, which takes into consideration reviews and ratings to faithfully recommend products to a user. Moreover, the results from the tables suggest that ratings alone do not reflect the satisfaction of users, but a combination with the review text will help to better capture users' sentiments about a product. We evaluated the zero-shot sentiment analysis model with the polarity values of the review text it gave. It was found to be satisfactory, hence it was adopted. In further work, we would like to consider different pre-trained models to compare to our current work.

## Discussion

We tested our LLMRS model and baseline with the query text, "HR program for managing employee records on the Windows platform". Figures 3 and 4 show that our model recommends a software with 591.79 ranking score as the best, the price for the software is $3,999; the implementation cost is $1,999.50 with maintenance cost of $399.90, compared to the baseline that recommended a product with an average rating of 5, which is the maximum rating, but with a price of $21,780, an implementation cost of $10,890, and a maintenance cost of $2,178. The baseline model of the best recommended software has only one review, which is not informative and rated as 5, while with our LLMRS model, the recommended software has more reviews, which together contribute to the ranking score, which is 591.79, implying that there were more than 100 reviews for the software, and most of the reviews were positive, giving evidence that from review text more information can be gathered to help decide on a purchase other than relying on ratings only, which in most situations are biased and give misleading information.

## Conclusion

We proposed an LLM-based zero-shot recommender system that encodes user reviews to make recommendations. A notable limitation of using user review information is that it requires a sizable number of reviews per product to rank a product. To test the efficacy of LLMRS, we performed experiments on the Amazon software product review dataset, which is enriched with user reviews. Results show LLMRS captures meaningful information from user reviews and uses it as an additional feature that contributes to ranking products. It overcomes the limitations of biased user reviews, such as a good review text recommending the product while giving it a low rating. The top recommendations are then selected based on the ranking score.

In e-commerce shopping, ranking-based recommendation systems are still widely used. These shopping websites contain an extensive number of product reviews. LLMRS has the potential to provide reliable recommendations using product review information without requiring additional computational sources. We expect LLMRS to improve recommendations for e-commerce shopping when applicable.

In future work, we could consider using a user's product reviews in the recommendation to help provide a personalized recommendation.